\documentclass[dvips, a4paper, 12pt]{article}

\usepackage{indentfirst}
\usepackage[english,francais]{babel}
\usepackage[latin1]{inputenc}

\usepackage{color}
\definecolor{oneblue}{rgb}{0,0.0,0.75}
\usepackage[colorlinks,
            urlcolor=oneblue,
            linkcolor=oneblue,
            citecolor=oneblue,
            bookmarksopen=false,
            pagebackref]{hyperref}

\usepackage[dvips,final]{graphicx}

\usepackage[sort,sectionbib]{natbib}

\usepackage{xspace} 
\usepackage{amsmath,amssymb,amsthm}

\usepackage{fancyhdr}
\def\R{\mathbb{R}}

\newcommand{\pd}[2]{\frac{\partial#1}{\partial#2}}
\newcommand{\od}[2]{\frac{d#1}{d#2}}

\newcommand{\abs}[1]{\left|#1\right|}
\newcommand{\norm}[1]{\left|\left|#1\right|\right|}

\def\og{\leavevmode\raise.3ex\hbox{$\scriptscriptstyle\langle\!\langle$~}}
\def\fg{\leavevmode\raise.3ex\hbox{~$\!\scriptscriptstyle\,\rangle\!\rangle$}}

\title{Linear theory of wave generation by a moving bottom}
\author{\href{http://www.cmla.ens-cachan.fr/\~dutykh}{Denys Dutykh%
\footnote{Centre de Mathématiques et de Leurs Applications, 
\'{E}cole Normale Supérieure de Cachan,
61, avenue du Président Wilson, 
94235 Cachan cedex, France}} \and
\href{http://www.cmla.ens-cachan.fr/\~dias}{Frédéric Dias\footnotemark[1]} \and
\href{}{Youen Kervella}\footnote{IFREMER, Laboratoire DYNECO/PHYSED, BP 70, 29280 Plouzan\'e, France}}
\date{}

\begin{document}

\selectlanguage{english}
\maketitle
\tableofcontents


\begin{abstract}
The computation of long wave propagation through the ocean obviously depends on the initial condition.
When the waves are generated by a moving bottom, a traditional approach consists in translating the ``frozen'' sea bed 
deformation to the free surface and propagating it. 
The present study shows the differences between the classical approach (passive generation) 
and the active generation where the bottom motion is included. The analytical solutions presented here exhibit 
some of the drawbacks of passive generation. 
The linearized solutions seem to be sufficient to consider the generation of water waves by a moving bottom.

\vskip 0.5\baselineskip

\selectlanguage{francais}
\noindent{\bf R\'esum\'e}
\vskip 0.5\baselineskip
\noindent
{\bf Théorie linéaire de génération de vagues par mouvement du fond.}
Les calculs de propagation d'ondes longues à travers l'océan doivent naturellement être alimentés par
la condition initiale. Le but de cette note est de montrer l'insuffisance de l'approche 
classique qui consiste à translater la déformation ``gelée'' du fond vers la surface libre et à la laisser se propager.
Un calcul analytique qui met en évidence les inconvénients de l'approche classique de génération passive
est présenté ici. Les solutions linéarisées semblent être bien adaptées pour traiter la génération
de vagues par mouvement du fond. 

\vskip 0.5\baselineskip
\noindent{\small{\it Mots-cl\'es~:} Vagues linéarisées, Mécanique des fluides, Problème de Cauchy-Poisson, Génération des tsunamis}
\end{abstract}

\section*{Version française abrégée}

Le problème de la génération des tsunamis est un sujet relativement récent. L'un des pionniers dans ce domaine fut Hammack
\cite{H73}. Le but de cette note est d'apporter une contribution à ce problème. La condition initiale utilisée dans les 
codes de propagation des tsunamis est souvent obtenue en translatant à la surface libre la déformation du fond suite à 
un tremblement de terre. Cette approche présente plusieurs inconvénients.
Tout d'abord la dynamique du processus de génération est négligée. Il est évident qu'un glissement lent ne produit pas
des vagues de même amplitude qu'un glissement rapide. Ensuite le champ des vitesses initiales est également négligé. Nous 
présentons dans cette note un simple modèle de génération de tsunamis (voir \cite{DD06} pour plus de détails).

Le problème des ondes de surface est tout d'abord linéarisé (\ref{lapl})--(\ref{kinsolb}) pour un fond qui a un mouvement
prescrit (génération active). Il est résolu par la méthode des transformées de Laplace en temps et de Fourier en espace. En supposant 
que le mouvement du fond est instantané, on obtient pour la déformation de la surface libre $\eta_i(x,y,t)$ l'expression
(\ref{instsol}). Les vitesses peuvent également être calculées. Le problème des ondes de surface est ensuite linéarisé
dans le cas où la déformation du fond est simplement translatée jusqu'à la surface libre (génération passive). Ce problème est différent
du précédent. En effet, puisque le fond reste immobile en tout temps, la condition cinématique au fond devient (\ref{kinsolbbis}).
Par ailleurs, la condition initiale sur la surface libre devient $\eta(x,y,0) = \zeta(x,y)$, où $\zeta(x,y)$ représente
la déformation permanente du fond. On obtient alors pour la déformation de la surface libre $\eta(x,y,t)$ l'expression
(\ref{genintsolbis}).

On compare ensuite les deux solutions (\ref{instsol}) et (\ref{genintsolbis}). Dans les deux cas on utilise la même
déformation du fond due à un tremblement de terre, qui est donnée par la solution d'Okada \cite{Ok92} pour une faille
finie rectangulaire située à une profondeur de $3$ km, de longueur $6$ km et de largeur $4$ km. Les autres paramètres
sont: module d'Young $=9.5$ GPa, coefficient de Poisson $=0.27$, glissement $=15$ m. La profondeur d'eau est $1$ km et
l'accélération de la gravité est $g = 10$ m/s$^2$. Pour ce type de déformation du fond, la vague initiale a la forme d'un $N$
et l'axe $y$ représente la direction privilégiée pour la propagation des ondes. La figure (\ref{fig:gauge}) montre le profil
des vagues mesuré à plusieurs endroits le long de la surface libre. La courbe en trait plein  
représente la solution dynamique tandis que la courbe en traits pointillés représente le scénario de génération passive.
Les amplitudes sont clairement plus grandes dans ce dernier cas. La figure (\ref{fig:differ}) montre la différence relative
(\ref{reldiff}) entre les deux solutions. Il y a deux différences essentielles entre les deux solutions: la génération
passive donne des amplitudes de vagues plus élevées et dans le cas de la génération active la colonne d'eau joue le 
rôle d'un filtre qui atténue les hautes fréquences grâce à la présence du cosinus hyperbolique au dénominateur. Les résultats 
dépendent naturellement de l'échelle de temps 
caractéristique de la déformation du fond. Dans le futur, nous allons également étudier l'effet des termes non-linéaires.

\selectlanguage{english}

\section{Introduction}
Tsunami generation is a relatively recent topic inspired for example by the pioneer work of Hammack \cite{H73}.
Since then, progress has been moderate. The present note provides a contribution to the development of this field of hydrodynamics.

The computation of long wave propagation across the ocean is a complicated task. 
The accuracy of the results depends on different factors such as the numerical method, 
the discretization error, the mathematical model error and others.
The error made in the initial condition cannot be corrected by the numerical method 
and will propagate in space and time. In our opinion it is important to construct an initial condition that is as accurate as possible.
Surprisingly there has been relatively little research in this field.

The initial condition is often constructed as follows. 
One takes coseismic deformations predicted by various models 
(\cite{Ok92} is presently used in many cases) and translates them to the
free surface. The velocity field is assumed to be zero. Then, a finite difference code 
computes the gravitational wave train induced by this free-surface disturbance.

This approach has several drawbacks. First of all, the dynamic 
character of the tsunami generation process is not taken into account. It is obvious from  physical
intuition (and confirmed by relatively simple computations) that slow slip does not produce 
waves of the same amplitude as fast bottom motion. So, usually, the initial wave amplitude is either under- or over-estimated 
depending on the time characteristics of the source.
Moreover the initial velocity field in the fluid due to the moving bottom is completely neglected. 
Our computations show that this is not necessarily true.

The present note sheds some light on these drawbacks. At the same
time the model studied in the present note can be considered as one of the simplest models for dynamic tsunami
generation. We refer to \cite{DD06} for more details.

\section{Linearized waves}
\subsection{Moving bottom solution}
Let us consider a three-dimensional fluid domain $\Omega$ bounded above by the free surface of the ocean $z=\eta(x,y,t)$
and below by the rigid ocean floor $z = -h + \zeta(x,y,t)$. The domain $\Omega$ is unbounded in the horizontal
directions $x$ and $y$, and can be written as $\Omega = \mathbb{R}^2\times\left[-h+\zeta(x,y,t),\eta(x,y,t)\right]$.
It is assumed that the
fluid is incompressible and the flow irrotational. The latter
implies the existence of a velocity potential $\phi(x,y,z,t)$ which
completely describes this flow. Initially the fluid is assumed to be at rest and the sea bottom to be 
horizontal ($z=-h$). Mathematically these conditions can be written 
in the form of initial conditions
$\phi(x,y,z,0) = 0$, $\eta(x,y,0) = 0$ and $\zeta(x,y,0) = 0$,\footnote{The last condition is not
an initial condition. We added it in order to have a flat bottom initially. In fact,
it is not required for the mathematical method.} which complete the formulation of the initial boundary 
value Cauchy-Poisson problem described below.
Thus, at time $t=0$, the free surface and the sea bottom are defined by $z=0$ and $z=-h$,
respectively. At time $t>0$ the bottom boundary moves in a
prescribed manner which is given by $z = -h + \zeta(x,y,t)$.
The displacement of the sea bottom is assumed to have all the properties
required to compute its Fourier transform in $x,y$ and its Laplace transform in $t$. 
The resulting deformation of the free surface
$z=\eta(x,y,t)$ must be found. 

Solving this problem is a difficult task due to the
nonlinearities and the a priori unknown free surface. In this study
we linearize the equations and the boundary conditions. 
The linearized problem in dimensional variables reads

\begin{equation}\label{lapl}
  \Delta \phi = 0, \qquad (x,y,z) \in \mathbb{R}^2\times[-h, 0],
\end{equation}
\begin{equation}\label{kinfreesurf}
  \pd{\phi}{z} = \pd{\eta}{t}, \qquad \pd{\phi}{t} + g\eta = 0, \qquad z = 0,
\end{equation}
\begin{equation}\label{kinsolb}
  \pd{\phi}{z} = \pd{\zeta}{t}, \qquad z = -h.
\end{equation}

Combining equations (\ref{kinfreesurf}) yields the single free-surface condition
\begin{equation}\label{singlefreesurf}
  \pd{^2\phi}{t^2} + g\pd{\phi}{z} = 0,
  \qquad z = 0.
\end{equation}

The problem (\ref{lapl}), (\ref{kinsolb}), (\ref{singlefreesurf}) can be solved by using the method of integral transforms. We apply
the Fourier transform in $(x,y)$ and the Laplace transform in time $t$.
For the combined Fourier and Laplace transforms, the notation $\overline{F}(k,\ell,s)$ is introduced.
After applying the transforms, equations (\ref{lapl}), (\ref{kinsolb}) and
(\ref{singlefreesurf}) become
\begin{equation}\label{lapltrans}
  \od{^2\overline{\phi}}{z^2} - (k^2+\ell^2)\overline{\phi} = 0,
\end{equation}
\begin{equation}\label{kinsolbtrans}
  \od{\overline{\phi}}{z}(k,\ell,-h,s) = s\overline{\zeta}(k,\ell,s),
\end{equation}
\begin{equation}\label{freesurftrans}
  s^2\overline{\phi}(k,\ell,0,s) + g\od{\overline{\phi}}{z} (k,\ell,0,s) = 0.
\end{equation}
The transformed free-surface elevation can be obtained from
(\ref{kinfreesurf}):
\begin{equation}\label{transfreesurf}
  \overline{\eta}(k,\ell,s) = -\frac{s}{g}\overline{\phi}(k,\ell,0,s).
\end{equation}
A general solution of equation (\ref{lapltrans}) is given by
\begin{equation}\label{gensollapl}
 \overline{\phi}(k,\ell,z,s) = A(k,\ell,s)\cosh(mz) + B(k,\ell,s)\sinh(mz),
\end{equation}
where $m = \sqrt{k^2+\ell^2}$. The functions $A(k,\ell,s)$ and $B(k,\ell,s)$
can be easily found from the boundary conditions (\ref{kinsolbtrans}) and
(\ref{freesurftrans}):
\begin{multline*}
  A(k,\ell,s) = -\frac{gs\overline{\zeta}(k,\ell,s)}{\cosh(mh)[s^2+gm\tanh(mh)]}, \\
  B(k,\ell,s)  = \frac{s^3\overline{\zeta}(k,\ell,s)}{m\cosh(mh)[s^2+gm\tanh(mh)]}.
\end{multline*}
From now on, the notation $\omega = \sqrt{g m\tanh(mh)}$ will be used.
Substituting the expressions for the functions $A$, $B$ in the general solution (\ref{gensollapl}) yields
$$
  \overline{\phi}(k,\ell,z,s) = -\frac{gs\overline{\zeta}(k,\ell,s)}
  {\cosh(mh)(s^2+\omega^2)}
  \left(\cosh(mz) - \frac{s^2}{gm}\sinh(mz)\right).
$$
From (\ref{transfreesurf}), the free-surface elevation becomes 
$\overline{\eta}(k,\ell,s) = s^2\overline{\zeta}(k,\ell,s)(s^2+\omega^2)^{-1}/\cosh(mh)$.

Now we assume that the sea bed deformation is instantaneous, i.e. $\zeta(x,y,t) = \zeta(x,y) H(t)$,
where $H(t)$ denotes the Heaviside step function\footnote{The Heaviside function has the property that 
it is equal to zero for $t\leq 0$. So, choosing 
this particular form for $\zeta$ satisfies automatically the condition $\zeta(x,y,0) = 0$.}. After 
some analytic computations one obtains the 
final integral formula for the free-surface elevation:
\begin{equation}\label{instsol}
  \eta_i(x,y,t)  = \frac{1}{(2\pi)^2}\int\!\!\!\int\limits_{\!\!\!\!\!\R^2}
  \frac{\widehat{\zeta}(k,\ell)e^{i(kx+\ell y)}}{\cosh(m h)}\cos\omega t\; dk d\ell,
\end{equation}
where $\widehat{\zeta}(k,\ell)$ is the Fourier transform of $\zeta(x,y)$. The velocity field due to the
moving bottom can also be computed \cite{DD06}. It cannot necessarily be neglected. 

\subsection{Passive generation}
In this case the initial condition is obtained by translating the sea bed deformation to the free-surface and 
the evolution of this system is computed. Next we give an analytic solution to this problem.
This solution is supposed to model what happens in the classical tsunami generation approach.

First of all, we have to make several modifications to the previous problem. 
Since the sea bed remains fixed at all time, the kinematic condition at the bottom becomes
\begin{equation}\label{kinsolbbis}
  \pd{\phi}{z} = 0, \qquad z = -h.
\end{equation}
The main difference with \S\,2.1 is the initial condition on free surface which 
becomes $\eta(x,y,0) = \zeta(x,y)$.

Again we apply the Fourier transform in the horizontal coordinates with the notation
$\widehat{F}(k,\ell)$. We do not apply
the Laplace transform because there is no substantial dynamics in this problem.
Equation (\ref{lapltrans}) is the same as before while (\ref{freesurftrans}) and (\ref{kinsolbbis}) 
become
\begin{equation}\label{freesurftransbis}
  \pd{^2\widehat{\phi}}{t^2}(k,\ell,0,t) + g\pd{\widehat{\phi}}{z} (k,\ell,0,t) = 0,
\end{equation}
\begin{equation}\label{inittrans}
 \pd{\widehat\phi}{z} = 0, \qquad z = -h.
\end{equation}

Since Laplace's equation still holds we have the same general solution (\ref{gensollapl}).
The relation between the functions $A(k,\ell,t)$ and $B(k,\ell,t)$
can be easily found from the boundary condition (\ref{inittrans}):
\begin{equation}\label{relationbis}
  B(k,\ell,t) = A(k,\ell,t)\tanh(mh).
\end{equation}
From equation (\ref{freesurftransbis}) and the initial conditions one finds $A(k,\ell,t)$
so that
\begin{equation}\label{phihatbis}
  \widehat{\phi}(k,\ell,z,t) = -\frac{g}{\omega} \widehat{\zeta}(k,\ell)
  \sin\omega t  \Bigl(\cosh(mz) + \tanh(mh) \sinh(mz)\Bigr).
\end{equation}

From the transformed dynamic condition $\widehat\phi_t+g\widehat\eta = 0$ at $z=0$, it is easy 
to find the Fourier transform of the free surface elevation
$$
  \widehat{\eta}(k,\ell,t) = \widehat{\zeta}(k,\ell)
  \cos\omega t.
$$

The inversion of the Fourier transform provides the simple integral solution
\begin{equation}\label{genintsolbis}
  \eta(x,y,t) = \frac{1}{(2\pi)^2}\int\!\!\!\int\limits_{\!\!\!\!\!\R^2}
  \widehat{\zeta}(k,\ell)  \cos\omega t \; e^{i(kx+\ell y)} dk d\ell.
\end{equation}

\subsection{Numerical computation}
We now compare the two solutions (\ref{instsol}) and (\ref{genintsolbis}). In both cases we use the same
sea bed deformation due to an earthquake which is given by Okada's solution \cite{Ok92} for a finite rectangular fault 
occurring at depth $3$ km with length $6$ km and width $4$ km. The other parameters are: Young's modulus $=9.5$ GPa, 
Poisson's ratio $=0.27$, dip angle $=13^\circ$, strike angle $=90^\circ$ and slip $=15$ m (dip-slip faulting). The water 
depth is $1$ km and the acceleration due to gravity is $g = 10$ m/s$^2$. For this particular sea bed deformation, the
initial wave has a $N-$shape and the $y-$axis is the preferred direction for wave propagation. 

All integrals were computed with a Filon-type numerical integration formula \cite{Fil28}, which
takes into account the oscillatory behaviour of the integrands.

\begin{figure}
	\centering
		\includegraphics[width=0.9\textwidth]{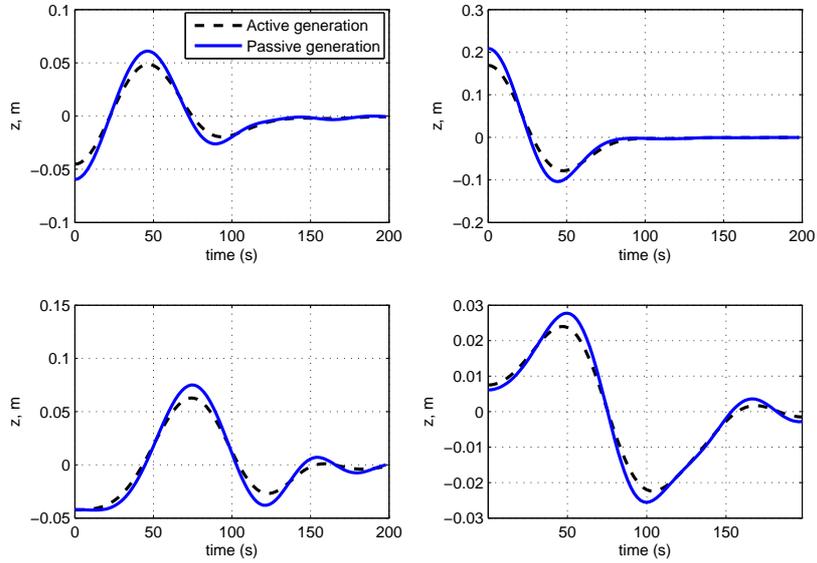}
	\caption{Wave profile $\eta(x,y,t)$ in km along several artificial tide gauges $(x,y)$ in km versus time $t$ in s}
	\label{fig:gauge}
\end{figure}

Figure (\ref{fig:gauge}) shows the wave profile measured at several locations along the free surface. The solid line 
represents the solution with instantaneous bottom deformation while the dashed line represents the passive 
wave generation scenario. The latter clearly exhibits higher wave amplitudes.

\begin{figure}
	\centering
		\includegraphics[width=0.85\textwidth]{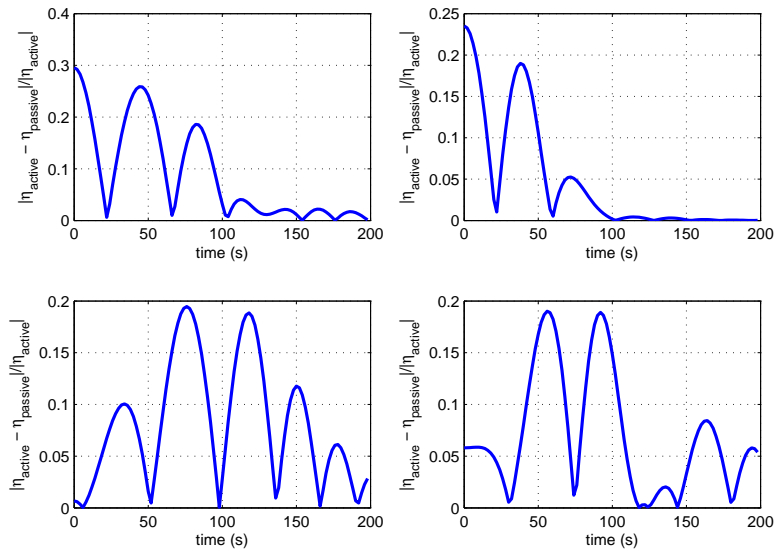}
	\caption{Relative difference between the solutions (\ref{instsol}) and (\ref{genintsolbis}) corresponding to
active and passive generation versus time}
	\label{fig:differ}
\end{figure}

Figure (\ref{fig:differ}) represents the relative difference between the two solutions which is defined by
\begin{equation}\label{reldiff}
  r(x,y,t) = \frac{\abs{\eta_i(x,y,t)-\eta(x,y,t)}}{\norm{\eta_i}_\infty}.
\end{equation}
Intuitively this quantity represents the deviation of the passive solution from that generated by a
moving bottom in units of the maximum amplitude of $\eta_i(x,y,t)$.

\section{Conclusions}
Looking at the analytic expressions for $\eta$ and
the numerical results gives some clear conclusions. Let us focus on two main differences which can be crucial
for accurate tsunami modelling.

First of all, the wave amplitudes obtained with the instantly moving bottom are lower than those generated by initial 
translation of the bottom motion (this statement follows from the inequality $\cosh mh \geq 1$ and a comparison between formulas 
(\ref{genintsolbis}) 
and (\ref{instsol})). The numerical experiment shows that this difference is typically of the order of $20\%$.

The second feature is more subtle. The water column has an effect of low-pass filter. 
It means that if the initial deformation contains high frequencies they will be attenuated in the moving bottom 
solution because of the hyperbolic cosine $\cosh(h\sqrt{k^2+\ell^2})$ in the denominator that grows exponentially with $m$.

Let us mention that if we prescribe a more realistic bottom motion as in \cite{DD06} for instance, the results 
will depend on the characteristic time of the sea-bed deformation. Even for very fast bottom motions, the generated wave amplitude 
will never reach the passive generation solution. For slow motions, the amplitude will be in general much smaller. 

Future studies will provide a more thorough development of this topic including the effect of different nonlinearities.

\bibliographystyle{plainnat}

\end{document}